\documentstyle[12pt,aasms4]{article}
\begin{document}

\title{Rapidly variable Fe K$\alpha$ line in NGC 4051}
\author{J.X. Wang$^1$, Y.Y. Zhou$^{1,3,4}$, H.G. 
          Xu$^2$, T.G. Wang$^{1,5}$}
\altaffiltext{1}{Center for Astrophysics,University of Science and 
     Technology of China, Hefei, Anhui, 230026, P. R. China; 
     jxw@mail.ustc.edu.cn}

\altaffiltext{2}{Institute for Space Astrophysics, Department of Physics, 
     Shanghai Jiao Tong University, Shanghai, 200030, P. R. China}  
   
\altaffiltext{3}{Beijing Astrophysics Center, Beijing, 100080, P. R. China}

\altaffiltext{4}{Laboratory of Cosmic Ray and High Energy Astrophysics, 
     Chinese Academy of Sciences, Beijing, 100039, P. R. China} 

\altaffiltext{5}{The Institute of Physical and Chemical Research (RIKEN),
     2-1, Horosawa, Wako, Saitama, 351-0198, Japan}

\authoremail{jxw@mail.ustc.edu.cn}

\begin{abstract}
We present a detailed analysis on the variability of the Fe K emission line in 
NGC 4051 using ASCA data. Through simple Gaussian line fits, we find not 
only obvious Fe K line variability with no significant difference in the 
X-ray continuum flux between two ASCA observations which were separated 
by $\sim$ 440 days,
but also rapid variability of Fe K line on time scales $\sim 10^4$ s within  
the second observation. During the second observation, 
the line is strong (EW = 733$^{+206}_{-219}$ eV) and broad 
($\sigma = 0.96^{+0.49}_{-0.35}$ keV) when the source is brightest, 
and become weaker (EW = 165$^{+87}_{-86}$ 
eV) and narrower ($\sigma<0.09$ keV) whilst the source is weakest. 
The equivalent width of Fe K line correlates 
positively with the continuum flux, which shows an opposite trend with 
another Seyfert 1 galaxy MCG --6-30-15.
\end{abstract}

 
\section{Introduction}

It has long been recognized that measurement of the profile
of the Fe K$\alpha$ fluorescence line found in many AGNs at $\sim$6.4 
 keV (Mushotzky 1995, Tanaka et al. 1995, Yaqoob et al. 1995, Fabian et al. 
1995, Nandra 
et al. 1997a and references therein) can provide an important tracer 
of matter in the vicinity of the postulated
supermassive black hole. 
Doppler and gravitational shifts would imprint 
characteristic signatures on the line profile
which map the geometric and dynamical distributions of
matter surround the black hole. Additional information concerning the 
geometry of matter in the active nucleus could in principle be derived 
by studying the rapid variability of the line profile, 
intensity and their relationship with the continuum variations. So far, 
rapid variability in Fe K line has been detected only in two Seyfert 
galaxies.   

Yaqoob el al. (1996) presented the evidence for rapid variability of 
the Fe K line profile in the narrow-line Seyfert galaxy NGC 7314, which is 
consistent with a diskline of constant equivalent width superposed on a 
constant flux narrow line (presumably from the torus). 
They found while the X-ray continuum flux varied by a factor of 2 on a 
time scale of hundreds of seconds, the emission in the red wing of the Fe K 
line below $\sim$6 keV responded to those variations on the time scales of 
less than $\sim3\times10^4$ s, and the response becomes slower and slower 
towards the line peak near 6.4 keV.  
Rapid variability of the Fe K line profile was also found in the bright
Seyfert 1 galaxy MCG --6-30-15 (Iwasawa et al. 1996). 
On time scales of less than $\sim10^4$ s, the variability pattern is the 
same as seen in NGC 7314. 
But on long time scales of about several 10$^4$ s, it shows a different 
behavior. When the source 
is bright, the Fe K line is weak and dominated by the narrow core, 
whilst during a deep minimum, 
a huge red tail appears. The intensity of broad Fe K line correlates
inversely with the continuum flux, so does the equivalent width.

In this Letter we present an evidence of rapid variability of Fe K 
line in another nearby (Z = 0.0023) low luminosity narrow-line 
Seyfert 1 galaxy NGC 4051, which is well known for its rapid variability
in the X-ray band. Nandra \& Pounds (1994) 
found an equivalent width of Fe K line $140\pm70$ eV from the Ginga 
data by assuming a single narrow line. A narrow Fe K emission line was also 
detected with an upper limit on the FWHM of $\sim460$ eV and equivalent 
width EW $\sim170$ eV during the first ASCA observation in 1993 (Mihara el 
al. 1994, hereafter M94) for this object. When the target was 
re-observed by ASCA in 1994 (Guainazzi el al. 1996, hereafter G96), the 
Fe K line was found to be stronger (EW = $350^{+170}_{-150}$ eV) and 
broad($\sigma>0.2 keV$).

\section{The ASCA data} 
NGC 4051 was observed twice by ASCA from 1993 April 25 22:30 to April 26 
21:30 and from 1994 July 6 15:28 to July 9 10:40. In this 
Letter, we 
concentrated on the Solid-state Imaging Spectrometer (SIS) data because 
of the better energy resolution it provides (Inoue 1993). The data 
reductions were performed with the ASCA standard software XSELECT 
according to the following criteria: satellite not passing 
through the South Atlantic Anomaly, geomagnetic cutoff rigidity 
greater than 6 GeVc$^{-1}$, and minimum elevation angle above Earth's 
limb of $10^{\rm o}$ and $20^{\rm o}$ for nighttime and daytime 
observations, respectively. 
Source counts were extracted from a circular area of radius 4' for the 
SIS0 and SIS1. Because of an error in satellite pointing during the first 
observation, the image fell into the dead region between two chips in 
SIS1 (M94). Therefore we only used SIS0 data from this observation 
for spectral analysis.
To avoid the complexity caused by the "warm absorber" and "soft excess" 
(M94, G96), we use the 
2.5--10.0 keV data for spectral analysis. Spectral fits were carried out 
using the XSPEC, and the background was taken from the blank sky data. 

\section{Variability of Iron K Line}
\subsection{Comparisons between two observations}
Time average spectra of Fe K$\alpha$ line profiles for the two observations 
were given by Nandra el al. (1997a) \& G96.
Due to the poor statistics, the line profile cannot be measured exactly, 
therefore we model the line with a single Gaussian function.
The underlying continuum is fitted with a single
power-law absorbed by the Galactic column 
density ($N_H\approx$1.3$\times$$10^{20} cm^{-2}$).  
We did not take into account of the possible Compton reflection
component here because of its small impact on the ASCA spectra.

Results of the single Gaussian fits are given in Table 1. 
For comparison, the results of diskline (Fabian el al. 1989, George 
\& Fabian 1991) fits can be found in Nandra el al. (1997a) \& G96.  
The variability of Fe K line is evident (EW changes from 166 eV to 330 eV 
and FWHM from 4500 km/s to 50000 km/s), however, there is no obvious 
variability in the photon index of X-ray continuum. 
Because of the pointing error of the satellite during the first 
observation, part of the photons fell into the gaps between the CCD 
chips. The continuum flux of Obs1 in Table 1 is only a 
lower limit. However, since the gaps occupy a fairly small
fraction of the selected area that includes our target ($\sim18\%$), the
continuum flux obtained cannot be biased too much. Thus we conclude
that there is no significant difference in the continuum flux
between two observations.

\subsection{Rapid FeK Line variability in the second observation}
During the two ASCA observations, NGC 4051 shows large amplitude X-ray 
continuum flux changes on time scales of $\sim$100 s. In order to see if 
there are short time scale variations of Fe K line within a single 
observation, we choose Obs2 which last longer time ($\sim$3 days) for 
analysis in detail.
Including a large flare in the beginning and a deep minimum near the end, 
the whole observation has been divided into five time 
intervals (from i1 to i5) with a similar exposure  time ($\sim$10 ks) as 
shown in Fig. 1 (different from G96). 
Among the five intervals,  i-2, i-3, i-4 have similar continuum fluxes.

Results of single Gaussian fittings are shown in Table 1 (We also 
give the results of diskline fits to the five intervals, 
respectively). The photon index
is quite similar between i2 to i4 at about $\Gamma\sim$1.80. 
A steeper continuum is suggested for i-1 and a flatter one for i-5. 
The single
Gaussian line dispersions and equivalent widths are also similar between
i2 to i4. 
A stronger and broader line is detected in i-1, and a weaker narrower one 
in i-5 (Fig. 2). 
In order to show the variability of Fe K line more clearly,
we present a contour plot of Gaussian line width versus
the equivalent width for i-1, i-5 and i-2+3+4 (a summed dataset of 
i-2,i-3,i-4)
obtained from the single Gaussian fits in Figure 3. 
For i-1, the equivalent width of Fe K line is 733 eV, much larger than 
that for the time average spectrum, and the Fe K line is much 
broader ($\sigma$ = 0.96 keV). While for i-5, the EW of Fe K line is only 
165 eV, and much narrower (only a narrow line has been 
detected, $\sigma<$0.09 keV). 
Due to the poorer statistics in i-5, there is concerning that 
a weak broad Fe K component might be un-noticeable.
To exclude this possibility, we add a broad Gaussian 
line (E$_{G} = 6.4$ keV, $\sigma = $0.46 keV, the same with the time 
average spectrum) to the fit. The upper limit for the EW of this component 
is 80 eV, far smaller than those in i-1 to i-4. These results clearly 
demonstrate the large variability of Fe K line during the second ASCA
observation. G96 also analyzed the Fe K line, but failed to detect 
the variability of Fe K line profiles and equivalent widths. 
We would like to point out here that the time sequence
selection used in G96, different to what we did in this paper, tends to 
smear out the line variations.

\section{Discussions}
\subsection{The X-ray continuum flux versus the equivalent width of Fe K
 line}  

We detect significant change not only in line equivalent width  but also in 
line profile through single Gaussian fitting to the two observations. From
1993.4.25 to 1994.7.6, Fe K line in NGC 4051 became broader and stronger, 
in spite of the fact the continuum flux remains almost constantly. 
This shows that the long time scale variability of Fe K line can be 
independent of the continuum variation. Such kind of behavior can be 
induced by the change in structure of the line emission region or the 
geometry of the hard X-ray source in the context of disk line model. The 
broader and stronger line emission during the 1994 epoch indicates that 
the average line emission region is much closer to the putative accretion 
disk. 
We also detect rapid variability ($\sim10^4$s) of Fe K line within the 
second observation.  
During the bright flare (i-1), the Fe K line is broad and strong, while 
during the deep minimum (i-5), Fe K line is narrower and weaker. 

Using the ASCA data of 39 AGNs, Nandra el al. (1997b) found a clear 
decrease in the strength of the Fe K$\alpha$ line with the increase 
in the luminosity (X-ray Baldwin effect, Iwasawa \& Taniguchi 1993), 
which was thought due to the fact that high-luminosity AGNs have high 
accretion rates, causing the accretion disk to become ionized (Matt et 
al. 1996). 
An inverse correlation between the Fe K line EW and the X-ray continuum flux 
was also found in MCG --6-30-15. 
The line EWs (sum of the broad and narrow component, calculated at the 
centroid energy 6.4 keV, see Table 2, Iwasawa et al. 1996) versus 
continuum flux is plotted here in Fig. 4b.
The case for the NGC 4051 shows that the situation is more complicated 
than this. We find an increase in the equivalent width with no apparent  
change in the continuum flux when the two ASCA observations are considered.
Moreover, an increase in the line EW with the increase X-ray continuum
flux (Fig. 4a), completely opposite to the Baldwin effect and the variability
in MCG --6-30-15, is found during the second ASCA observation. It 
indicates a complicate mechanism should be considered. 

\subsection{rapid variability on time scale $\sim10^4$s}

It is well known that the X-ray source in many AGNs is highly variable
on short time scales $\sim$100 s. As Fe K line is produced via fluorescent 
process, its strength and profile should response to the continuum 
variability. 
Obviously, the 
measurement of the temporal response is the best way to map the matter 
distribution in innermost regions around the black hole. In principle, 
the distance from the X-ray continuum source to the line emission region 
could be determined through observing reverberation effect (Reynolds et 
al. 1998). With the Fe K line profile, we could derive the mass of the 
black hole in the heart of nucleus. The characteristic time scales on 
which reverberation effects occur is the light crossing time of one 
gravitational radius, which is $\sim$50 s for a 10$^7$ M$_{\sun}$ black 
hole. But, all current 
available instruments are not able to measure such a fast time response.
Even for a bright Seyfert galaxy, ASCA can only obtain 1$\sim$2 Fe K line 
photons in 100 s, too few to define the line flux and profile.
However, the Fe K line variability of time scales $\sim10^4$ s can provide 
constraints on the long time scale ($\sim10^4$ s) variation of the X-ray
source or the line emission region.
So far, such variability has been detected
in only three Seyfert galaxies, NGC 7314, MCG --6-30-15 and NGC 4051. For 
NGC 7314,
a simple disk-plus-torus model could explain its rapid variability 
fairly well. But for MCG --6-30-15 and NGC 4051, the situation is more 
complicated.

According to the disk model, the fluorescent iron lines is produced from the 
hard X-ray irradiation of the disk composed of cold gas around a black hole.
When the X-ray source is bright, iron will start to be ionized, and in the 
intermediate ionization states, the resonance scattering can cause a 
reduction in the line flux (Matt et al. 1996). It is a possible reason
causing the anti-correlation of the EW of broad components to the X-ray 
continuum flux in MCG --6-30-15. We consider that the ionized disk is also a 
possible reason for the positive correlation in NGC 4051 because a factor 
$\sim$2 larger EW for FeXXV than cold iron could occur (Matt et al. 1996).
Though not well constrained, the diskline energies for i1 to i5 in Table 
1. show an trend to correlate with the continuum flux, which gives some 
support to this opinion.

There is also another possible explanation to the the rapid variability 
of Fe K line.
Fabian (1997) suggested that
the X-ray continuum might be generated by magnetic flares above the 
accretion disk. At any given time there are only a few flares otherwise
the rapid variability will be average out. When the dominated flares move
around on the disk, changes in the Fe K line profile would be seen. Such
model could explain the odd variability of Fe K line in MCG --6-30-15 well.
We think it is also appropriate for the rapid variability in NGC 4051
discovered in 
this paper. For the sequence i-5, the Fe K line profile is similar to the 
bright flare in MCG --6-30-15. The small FWHM might due to a dominate flare 
located far away from the central black hole, or a 
succession of flares 
on the approaching side of the disk. The broad line during i-1 might due 
to a bright dominate flare very close ($\sim$6$R_g$) to the black hole, and 
the large EW might due to a overabundance of iron.

\acknowledgments
This work is supported by Chinese National Natural Science Foundation, 
PanDeng Project and Foundation of Ministry of Education.
The authors would also like to thank S.A. Huang for great help with
ASCA data reduction.

\newpage 
\begin{figure}
\input psfig.sty
\psfig{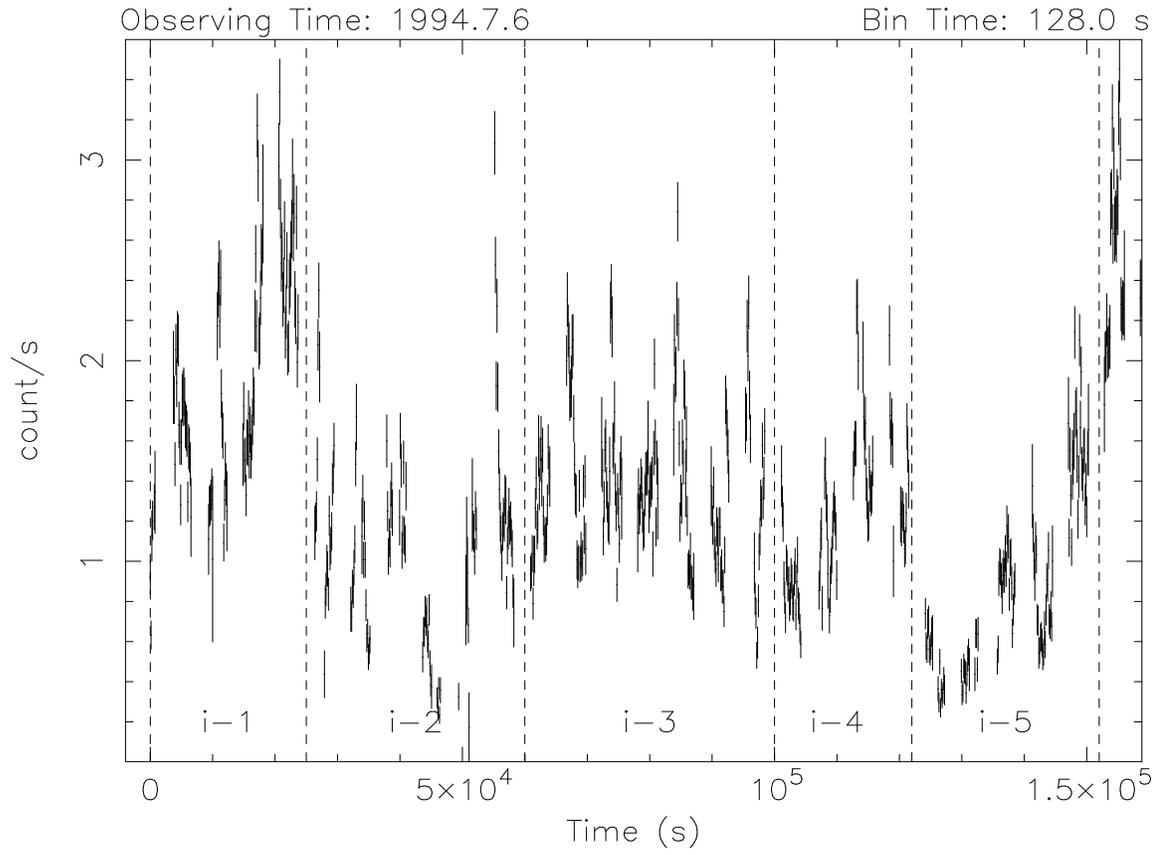}      
\figcaption[f1.ps]{The 0.4--10.0 keV light curve from the SIS0 for the 
second ASCA observation, binned at 128 s. Time-intervals used in the 
section 3.2 are also indicated here. \label{fig-1}}
\end{figure}

\newpage
\begin{figure}
\input psfig.sty
\psfig{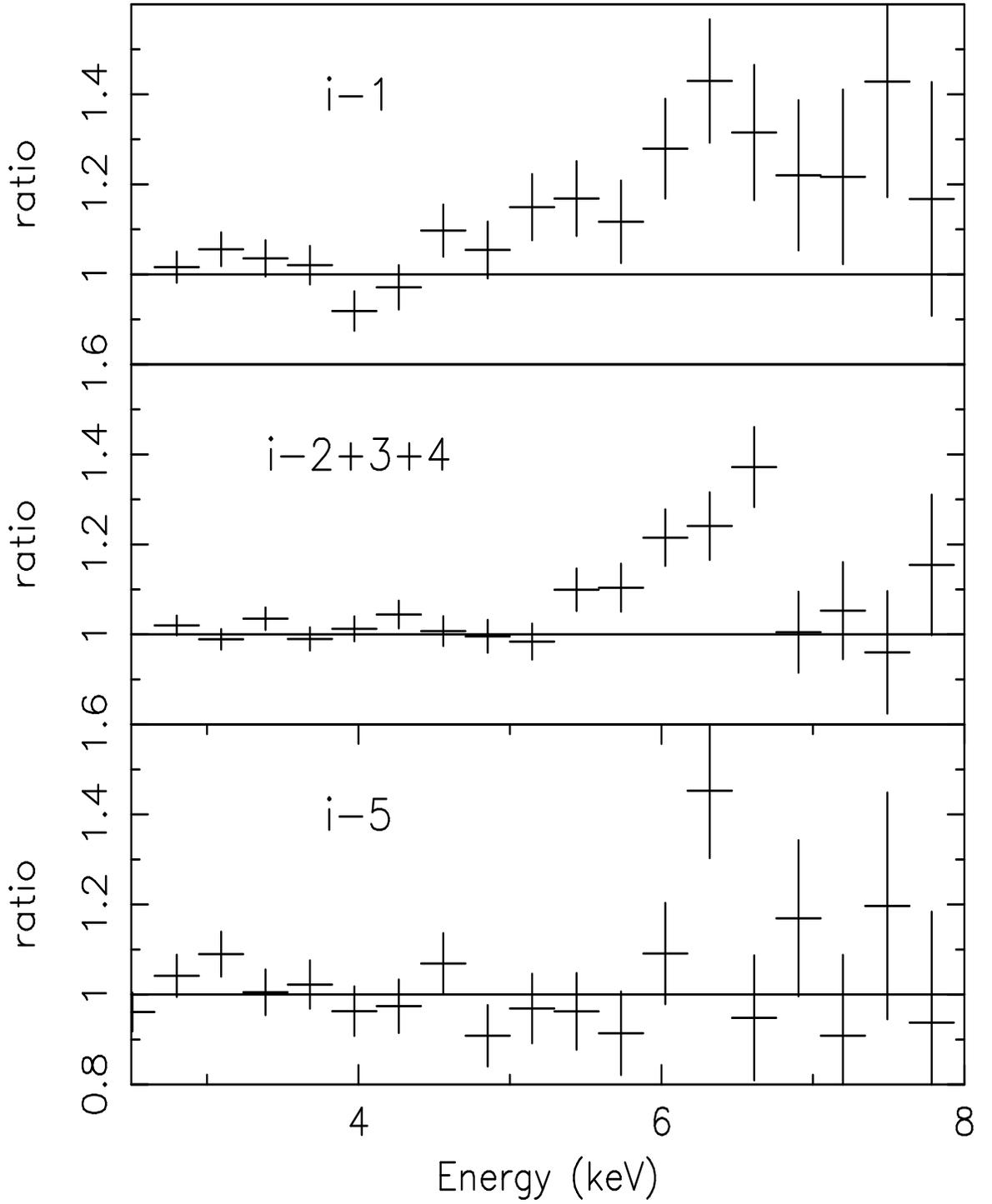}
\figcaption[f2.ps]{Ratios of the data to an absorbed power-law model fit to 
the data in the 2.5 to 4.5 keV and 7.0 to 10.0 keV bands to illustrate the 
Fe K line profile for the SIS in the time intervals i-1,i-2+3+4,i-5, 
which are indicated in Fig. 1. \label{fig-2}}  
\end{figure}

\newpage
\begin{figure}
\input psfig.sty
\psfig{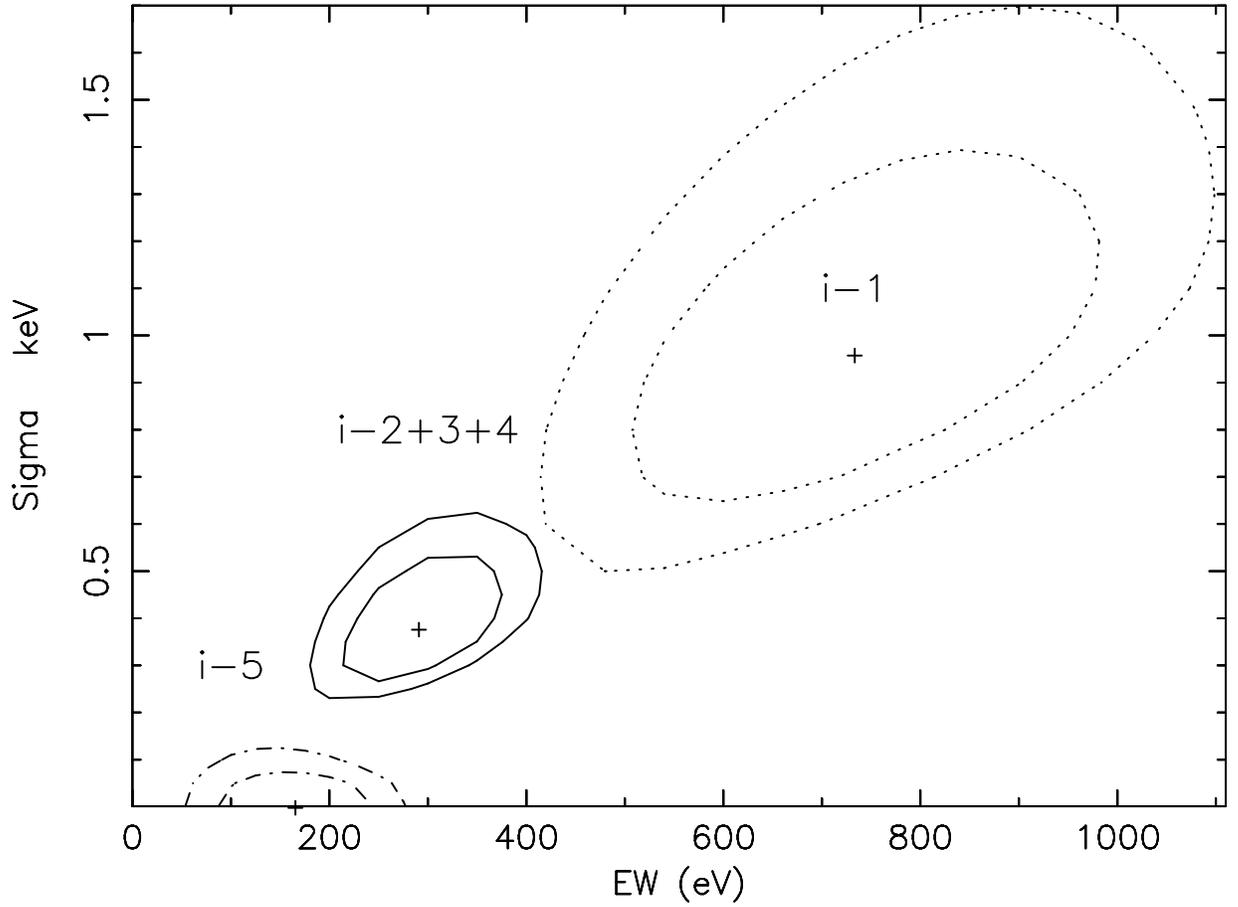}
\figcaption[f3.ps]{68 and 90 per cent contours in the line width and the 
line EW for two degrees of freedom from the single Gaussian fit to the 
spectra in the time-intervals i-1,i-2+3+4,i-5. \label{fig-3}}
\end{figure}

\newpage
\begin{figure}
\input psfig.sty
\psfig{figure=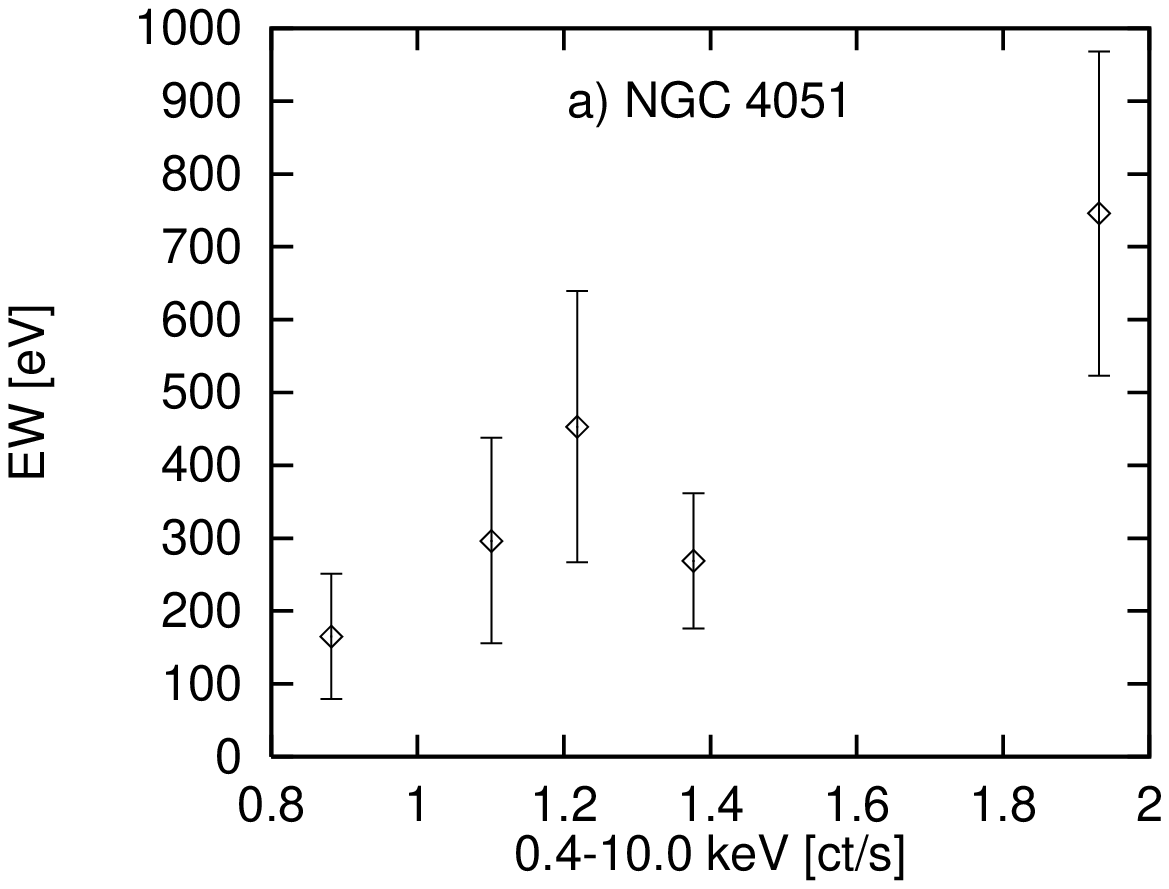,width=14.2cm,height=8.0cm,angle=-90}
\psfig{figure=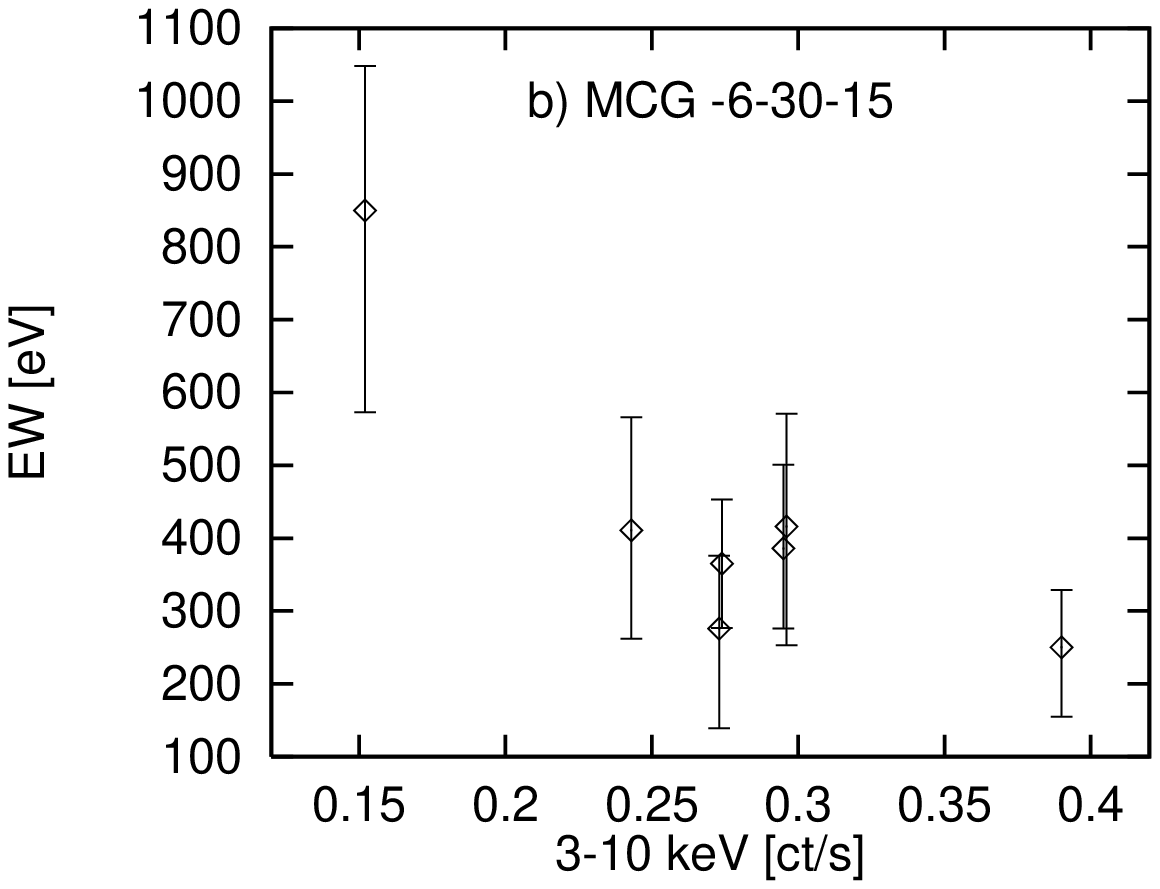,width=14.2cm,height=8.0cm,angle=-90}
\figcaption[f4a.ps,f4b.ps]{Plot of Fe K line EWs versus continuum flux 
for: a) NGC 4051 and b)MCG --6-30-15. \label{fig-4}}
\end{figure}

\newpage

\begin{table}
\caption[]{Single Gaussian fits to the two ASCA observations 
and the five datasets of Obs2} 
\begin{tabular}{ccccccccc}\hline \hline

... & $\Gamma^a$ & E$_{G}$ (keV) & $\sigma$ (keV) & EW (eV)$^c$ & 
$\chi^2/dof$ & flux$^b$ & E$_{D}^d$ (keV) & $R_i/R_g^d$

\\ \hline
Obs1 & $1.94^{+0.12}_{-0.11}$ & $6.49^{+0.05}_{-0.16}$ &  
$0.04^{+0.14}_{-0.04}$ & $162^{+72}_{-72}$ & 148/127 & $>2.15$\\ 
%& $6.49^{+0.05}_{-0.16}$ & $6.49^{+0.05}_{-0.16}$\\ 

Obs2 & $1.83^{+0.05}_{-0.06}$ & $6.35^{+0.14}_{-0.14}$ & 
$0.46^{+0.19}_{-0.14}$ & $330^{+68}_{-68}$ & 308/335 & 2.42\\ 
%& $6.41^{+0.09}_{-0.09}$ & $15.^{+70.}_{-4.}$\\ 

i-1 & $2.04^{+0.13}_{-0.13}$ & $6.45^{+0.44}_{-0.44}$ &
$0.96^{+0.49}_{-0.35}$ & $733^{+206}_{-219}$ & 184/190 & ... 
& $6.74^{+0.30}_{-0.40}$ & $6.3^{+2.4}_{-0.3}$\\

i-2 & $1.71^{+0.14}_{-0.13}$ & $6.02^{+0.28}_{-0.38}$ &
$0.44^{+0.21}_{-0.18}$ & $296^{+142}_{-140}$ & 174/172 & ... 
& $6.49^{+0.11}_{-0.11}$ & $13.^{+11.}_{-6.}$\\

i-3 & $1.83^{+0.10}_{-0.10}$ & $6.47^{+0.19}_{-0.13}$ &
$0.21^{+0.34}_{-0.13}$ & $269^{+ 93}_{- 93}$ & 228/227 & ... 
& $6.49^{+0.11}_{-0.11}$ & $13.^{+11.}_{-6.}$\\

i-4 & $1.87^{+0.16}_{-0.17}$ & $6.18^{+0.31}_{-0.29}$ &
$0.46^{+0.94}_{-0.27}$ & $453^{+186}_{-186}$ & 119/138 & ... 
& $6.61^{+0.16}_{-0.15}$ & $7.0^{+4.1}_{-0.9}$\\

i-5 & $1.51^{+0.15}_{-0.15}$ & $6.39^{+0.05}_{-0.05}$ &
$0.00^{+0.09}_{-0.00}$ & $165^{+ 87}_{- 86}$ & 152/149 & ... 
& $6.40^{+0.07}_{-0.09}$ & $999^{+1}_{-787}$\\

\hline \end{tabular}
$^a$Best fitting value when data in 4.5-7.0 keV region are
excluded\\
$^b\times10^{-11} $erg.cm$^{-2}$.s$^{-1}$ (2-10 keV)\\
$^c$calculated at the centroid energy 6.4 keV\\  
$^d$Results of diskline models fits with disk emissivity index $q$
fixed at -3.0, the outer disk radii $R_o$ at 1000$R_g$ ($R_g = GM/c^2$) and 
the inclination of the disk at 25$^{\rm o}$ (G96)\\
\end{table}       

\end{document}